\documentclass[fleqn,10pt]{wlscirep}
\usepackage{pdfpages}
\newcommand*\mystrut[1]{\vrule width0pt height0pt depth#1\relax}

\title{The Impact of Environmental Fluctuations on Evolutionary Fitness Functions
}

\author[1,*]{Anna Melbinger}
\author[1] {Massimo Vergassola}
\affil{University of California San Diego, Department~of Physics, 9500~Gilman Drive, La Jolla, CA 92093}

\affil[*]{amelbinger@ucsd.edu}

\begin{abstract}
The concept of fitness as a measure for a species's success in natural selection is central to the theory of evolution. We here investigate how reproduction rates which are not constant but vary in response to environmental fluctuations, influence a species' prosperity and thereby its fitness.  Interestingly, we find that not only larger growth rates but also reduced sensitivities to environmental changes substantially increase the fitness. Thereby, depending on the noise level of the environment, it might be an evolutionary successful strategy to minimize this sensitivity  rather than to optimize the reproduction speed. Also for neutral evolution, where species with exactly the same properties compete, variability in the growth rates plays a crucial role. The time for one species to fixate is strongly reduced in the presence of environmental noise. Hence, environmental fluctuations  constitute a possible explanation for effective population sizes inferred from genetic data that often are much smaller than the census population size.
\end{abstract}
\begin{document}

\flushbottom
\maketitle
%
%
\section*{Introduction}

Since its formulation by Darwin and Wallace, the theory of evolution and its explanation for the ongoing development of different species became a paradigm of modern biology~\cite{Darwin,Wallace:1858}. Herbert Spencer's famous expression ``survival of the fittest''~\cite{Spencer} provides an appealing and  concise summary of the concept of natural selection. However, it leaves aside one of the most complex  yet important aspects of evolutionary theory, namely identifying the factors determining the fitness of a species~\cite{Metz:1992,Ariew:2004}\,: fittest individuals are by definition the ones which prevail but the reasons facilitating their survival are not obvious. Even leaving aside the difficulties arising due to the genotype-phenotype mapping~\cite{Visser:2014}, it is far from trivial to identify a species' fitness function and its dependence on measurable ecological quantities.

Examples of determinants for evolutionary fitness are reproduction-related quantities like birth rate, viability, number of offspring and span of fertility. All of those directly influence the amount of genes that an individual will transmit to the future population (either carried by the individual itself or its offspring). Importantly, all those factors depend on the specific environment in a species' habitat. This fact is strongly related to the concept of niches:  in each niche a different species has potentially the largest fitness and outcompetes less adapted ones. Other ecological factors like population structure and composition also have bearing on this issue. Therefore, traditional fitness concepts solely based on growth rate and viability were extended by frequency-dependent~\cite{MaynardSmith:1973,Hofbauer} or inclusive fitness approaches~\cite{Hamilton:1964}. 

Environmental conditions are not constant but vary on almost every time scale and pattern. 
Whole niches change with time and space~\cite{Pearman:2007} but also within a well-defined niche constant environmental conditions seem to be the exception rather than the norm. For instance, the availability of different nutrients, the presence of detrimental substances or other external factors like temperature, all strongly  influence reproduction/survival and occur on a broad range of time scales~\cite{Mustonen:2007}. The relevance of environmental fluctuations for evolutionary dynamics was demonstrated in many different contexts, e.g. general consequences of environmental noise on growth and extinction~\cite{Levins:1968,Lewontin:1969, Gillespie:1973b,Takahata:1975,Levin:1984,Frank:1990,May:1973,Haccou:1994,Yoshimura:1996,Orr:2007,Chevin:2010,Cvijovic:2015,Desponds:2015}, more specific scenarios like the influence of environmental noise on evolutionary game theory or predator-prey models~\cite{Dobramysl:2013, Ashcroft:2014}, the invasion dynamics of new species~\cite{Neubert:2000}, its interplay with phenotypic variations~\cite{Riviore:2014}, phenotypic plascitity~\cite{DeWitt:2004} or role environmental tolerance~\cite{Lynch:1987}. Evolutionary strategies to actively cope with variable environmental conditions  like phenotypic heterogeneity or bet-hedging have been extensively studied as well~\cite{Schaffer:1974,Kussell:2005b, Kussell:2005,Acer:2008, Beaumont:2009,Pintu:2014}.

The scope of this paper is to quantitatively understand the impact of fluctuating reproduction rates on evolutionary dynamics. Specifically, the interplay of such dynamics  with demographic fluctuations was not fully elucidated yet. The latter becomes especially important as the crossover between selection-driven and fluctuation-driven evolution is a major focus of modern research on evolutionary dynamics~\cite{KimuraOhta,Ewens,Nowak:2004,Cremer:2009}. Environmental fluctuations potentially influence both neutral and selection driven evolution rendering a proper understanding essential  to grasp the dynamics. Here, we investigate this issue by combining analytical calculations and stochastic simulations.  Thereby, we show that an individual's sensitivity to environmental changes contributes substantially to its fitness: a reduced sensitivity increases the fitness and may compensate for large disadvantages in the average reproduction rate. We also find that fluctuating environments influence neutral evolution where they can cause much quicker fixation times than 
na\"ively expected. As we explain in detail in the following, this finding has interesting consequences for the interpretation of the effective population size which is typical characteristic to quantify randomness in an evolutionary process. Finally, we show that our results hold not only for very quickly fluctuating environment but also for switching rates up to the time scale of reproduction.

\section*{Results}

\begin{figure}
\includegraphics[width=\textwidth]{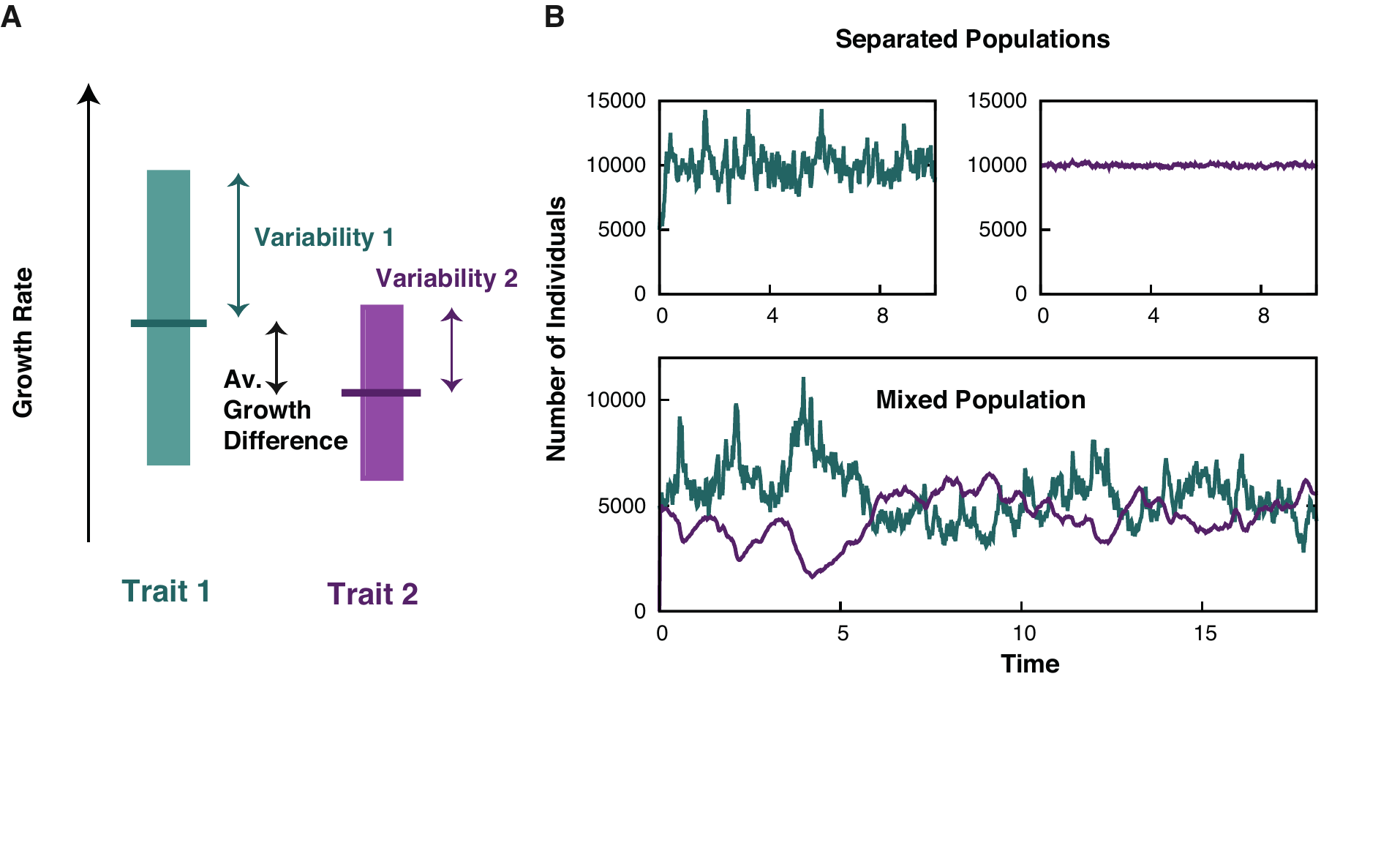}
\caption{{\bf A} Illustration of two species with different growth rates and sensitivities to the environmental fluctuations. The mean growth rates are indicated by horizontal lines and the variabilities by the length of the colored bars. {\bf B} Typical trajectories for two species with the same growth rates $\nu_1=\nu_2=10$ but different sensitivities to the environment ($\sigma_1=0.5$ and $\sigma_2=0$, as defined in the text), shown in green and violet, respectively. In the upper two panels the time evolution of two separate non-competing populations is shown. Due to their equivalent average growth rate, both reach the same carrying capacity but fluctuations are substantially stronger for species {\bf 1}. In the lower panel the evolution of both species in the same habitat is shown. Other parameters defined in the main text are the death rate $\gamma=1$ and the carrying capacity $K=1000$. \label{fig:cartoon}}
\end{figure}
To understand the impact of variable environmental conditions, we first consider an extension of a model introduced by May~\cite{May:1973}, which is an evolutionary process based on fluctuating  birth rates. Different species are defined by their specific traits which influence both their average reproduction rate as well as their sensitivity on environmental changes.The model assumes that populations grow logistically, i.e. the population grows exponentially if the total population size is small but reaches a finite maximal size after a while which is set by
limited resources. In contrast to standard logistic or Verhulst dynamics we here consider growth rates which are not necessarily constant but may fluctuate due to environmental changes. Mathematically such a scenario can be modelled by decoupled birth-death dynamics for each trait, $S$, with noise stemming from environmental changes and demographic fluctuations (the latter were not considered in Ref.~\cite{May:1973}). The dynamics is described by the following stochastic differential equations for the total number of individuals, $N_S$, of type $S$:
\begin{align}
\dot N_S=\underbrace{\left(\nu_S-\gamma\!\frac{N}{K}\right)N_S}_{\text{logistic growth}}+\underbrace{\mystrut{2.1ex}N_S\sigma_S\xi_S}_\text{env. noise}+\underbrace{\mystrut{2ex}\sqrt{\!N_S\left(\nu_S+
\gamma\!\frac{N}{K}\right)}\mu_S}_\text{demograph. noise}.
\label{eq:Langevin}
\end{align}
The first term is purely deterministic and accounts for reproduction and death events according to standard logistic growth~\cite{Verhulst, Hastings:1997}. In more detail, an individual of type $S$ reproduces at a growth rate, $\nu_S$ while the death rates are assumed to be  identical for all traits. Population growth is bounded and deaths rates increase with the total population size  $N\!=\!\sum_S N_S$ where $N_S$ is the number of $S$-type individuals. This  may account for density-dependent ecological factors such as limited resources or metabolic waste products accumulating at high population sizes. For specificity, we choose $\gamma N/K$ as functional form where $K$ is the carrying capacity scaling the maximal number of individuals and $1/\gamma$ sets the timescale of  death. 

Let us now consider the role of environmental noise. We assume that the environment directly acts on the reproduction rates as illustrated in Fig.~\ref{fig:cartoon}~{\bf A}. Thereby, variable environmental conditions can be modeled by fluctuating birth rates:
\begin{align}
\nu_S\rightarrow\nu_S+\sigma_S\xi_S\,,
\end{align}
where $\xi_S$ is  $\delta$-correlated white noise with $\langle \xi_S(t)\xi_{S}(t')\rangle=\delta(t-t')$ and the standard deviation (STD) $\sigma_S$ is the strength of that noise. Note that $\sigma_S$ is affected both by the actual noise level of the environment and the traits' sensitivity on the environment. Environmental noise acts on the birth rate and appears in Eq.~\eqref{eq:Langevin} multiplied by the number of individuals, $N_S$,  i.e. it is linearly multiplicative, which will be crucial for the results presented below.
 Beside environmental noise, also demographic fluctuations arising from the stochastic nature of the birth-death dynamics are present. Such fluctuations are more pronounced for smaller populations since the impact of a random event on the average is larger then. They lead to a phenomenon called `random drift' responsible for neutral evolution which causes extinction events even without selection as driving force. According to standard formulations, such demographic fluctuations here yield the term $\sqrt{N_S(\nu_S\!+\gamma N/K)}\mu_S$, where $\mu_S$ is $\delta$-correlated noise, $\langle \mu_S(t)\mu_{S}(t')\rangle\!=\!\delta(t-t')$, with a variance given by the sum of reaction rates~\cite{Gillespie}. We shall carry out further analysis for only  two different traits $S\in\{\text{\bf1,2}\}$ but generalizations to more traits are straightforward. 
 
 \subsection*{Noise Correlation and Fokker-Planck Description}
 
 As it will turn out, the correlation level of the noise is crucial for some important features of the model. Environmental noise acting on the growth rate can influence several species at the same time or act independently on each species. To capture such different noise correlation levels we introduce a correlation parameter, $\epsilon=\langle\xi_1\xi_2\rangle$. For instance, if several species feed from the same nutrients whose abundance fluctuates the noise of the growth rate of those species fluctuates with the same pattern (not necessarily the same amplitude). In this situation noise is perfectly correlated, i.e. $\epsilon=1$. But also other situations are reasonable: If both species' growth rates depend on different external variables, for example both feed from different carbon sources, noise is uncorrelated, $\epsilon=0$. Also anticorrelated noise is potentially possible, $\epsilon=-1$. This means that noise increases the growth rate of one species while it decreases the growth rate of another species. This might happen if one species metabolizes a substance which is poisonous for the other species. As all those scenarios are possible, we keep our analyses general by employing the correlation parameter, $\epsilon$, and discuss scenarios where a particular choice of the noise correlation changes the evolutionary outcome.
 
To analyze the evolutionary dynamics and its dependence on both environmental and demographic noise, it is useful to study  the Fokker-Planck equation (FPE) associated to Eq.~\eqref{eq:Langevin}. This equation cannot only be used to derive crucial quantities for the evolutionary process like fixation probabilities and times  but also offers the possibility to distinguish the contributions of Darwinian fitness and neutral evolution as we are going to explain the the following section.  In the remain of this section, the FPE for the relative abundance,$x=\frac{N_1}{N_1+N_2}$, will be derived which might be skipped by a mathematically less interested reader.
To do so, first the Langevin equations~\eqref{eq:Langevin} have to be transformed to a FPE in the variables $N_1$ and $N_2$. The procedure is straightforward but the correlation level of the noise has to be considered correctly~\cite{Gillespie:1996}. This leads to the following FPE also depending on the correlation parameter $\epsilon$,
 \begin{align}
 &\frac{\partial P(N_1,N_2,t)}{\partial t}=\epsilon \partial^2_{1,2}\sigma_1\sigma_2N_1N_2P - \sum_{i}\partial_{i}\left[\left(\nu_i-\gamma\frac{N}{K}\right)N_i P\right]+\frac{1}{2}\sum_i
 \partial^2_{i}\left\{ \left[(N_i\sigma_i)^2+ N_i\left(\nu_i+\gamma\frac{N}{K}\right)\right]P\right\}\,,
 \end{align}
 where $\partial_i\equiv \partial_{N_i}$. To uncover the influence of environmental noise on the evolutionary dynamics, the relative abundances seem the natural observables. Therefore, we change variables to the fraction $x=\frac{N_1}{N_1+N_2}$ and the total number of individuals $N=N_1+N_2$. 
The FPE for $x$ and $N$ can be simplified exploiting  the fact that the timescale of selection, $s=\nu_1-\nu_2$, is much slower than the timescale for population growth $\nu_1x+\nu_2(1-x)$. Therefore, we integrate over the total population size $N$, considering the FPE for the marginal distribution $P(x)=\int_0^\infty P(x,N)\,dN$, and employ $N\gg1$, see Supporting Material (SM). The resulting one-dimensional FPE reads\,:
 \begin{align}
 &\frac{\partial P(x,t)}{\partial t} = \partial_x\left\{ \left[-s\!-\!\sigma_2^2(1\!-\!x)\!+\!\sigma_1^2x\!+\!\epsilon\sigma_1\sigma_2\left(1-2x\right)\right]Q\right\}+\!\partial^2_x\!\left\{\!\left[\frac{\sigma_1^2\!-\!2\epsilon\sigma_1\sigma_2\!+\!\sigma_2^2}{2}x(1\!-\!x)\!+\!\!\frac{\gamma}{K}\right]\! Q\right\}\!\!\equiv\! {\cal L}P(x,t), \label{eq:FP}
\end{align}
where $Q\equiv x(1-x)P(x,t)$ and the last equality defines the Fokker-Planck operator ${\cal L}$ needed in the following. For $\sigma_1\!=\!\sigma_2=0$, the drift term reduces to the well-know expression $-s\partial_x x(1-x)P(x,t)$ favoring the trait with the higher growth rate~\cite{Kimura}. In Ref.~\cite{Takahata:1975} a similar FPE was derived for the special case $\sigma_1=\sigma_2$.  Note that, contrary to the analysis performed there, our model was first formulated considering the joint  evolution of the total number of individuals and their relative abundances and  then
Eq.~\eqref{eq:FP} for the relative abundances was derived by marginalization. That is crucial to fully capture the effects of noise.

\subsection*{Sensitivity to Environmental Changes as an Evolutionary Disadvantage}
With the one-dimensional FPE at hand, the evolutionary dynamics and especially the impact of environmental noise can be investigated. The equation has two terms: The first one proportional to $\partial_x$ describes the  drift due to selection on fitness differences and therefore corresponds to Darwinian evolution. Analyzing it gives us first insights about how the fitness depends on the sensitivity to environmental changes and which species is favored by selection. However, to fully understand the evolutionary outcome also the second term proportional to $\partial_x^2$ has to be considered. This term describes the impact of the random drift. For example if this term is much larger than the first one evolution is completely neutral and species outcompete other ones only by chance.

 To grasp the consequences of different sensitivities to environmental changes, we first discuss the case of {\it distinct environmental sensitivities}, defined by $\sigma_1=\Delta$ and $\sigma_2=0$, i.e. only the reproduction rate of the first trait depends on the environment. In Fig.~\ref{fig:cartoon}~{\bf B} we show typical trajectories for such a scenario. There, we consider two species with the same average reproduction rate, $s=\nu_1-\nu_2=0$ but different sensitivities $\sigma_1=\Delta=0.5$ (green) and $\sigma_2=0$ (violet). In the upper two panels the time evolution is shown when both species live in separate environments while in the lower panel their coexistence is considered. The question we are tackling in the following is whether or not there is an evolutionary force between those two species, in spite of their growth to the same total population size when living separated. The drift term in Eq.~\eqref{eq:FP} which is  proportional to $\alpha(x)=(s-\Delta^2 x)x(1-x)$ is key to answer that question. One can easily derive the mean field dynamics or deterministic limit from it,
\begin{align}
\partial_t \langle x\rangle=\left[s-\Delta^2\langle x\rangle\right]\langle x\rangle\left[1-\langle x\rangle\right]\,,
\end{align}
where $\langle x\rangle$ is the mean fraction of species  {\bf 1} in the population.
 If $s<0$, i.e. the second trait  with a smaller variability in its birth rate is also faster in reproducing, the evolutionary dynamics does not change qualitatively compared to $\Delta=0$. Conversely, if $s>0$, the situation changes dramatically\,:   the growth rate favors trait {\bf1} while the variability term favors trait {\bf2}. This leads to a stable fixed point $x^*=\frac{s}{\Delta^2}$ for $s<\Delta^2$ (for $s>\Delta^2$ variability is not sufficient to prevent extinction of trait {\bf2}). Such a dynamics can be interpreted as a frequency-dependent fitness function. However, the frequency-dependence arises here from environmental noise and not from 
a pay-off matrix as in standard evolutionary game theory~\cite{Maynard}. 
\begin{figure}[t]
\centering
\includegraphics[width=\columnwidth]{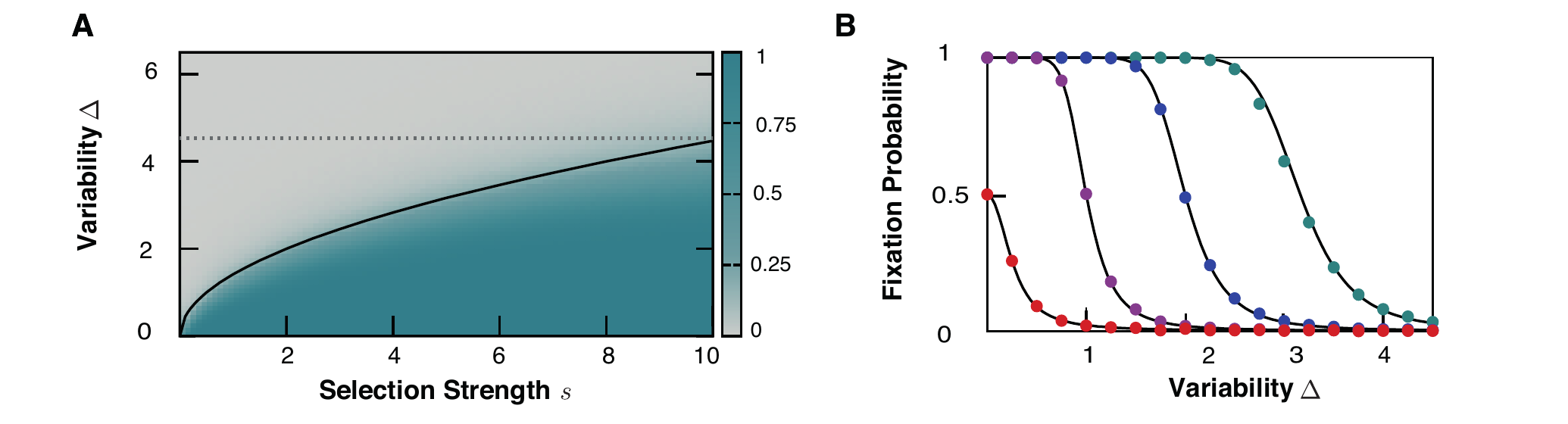}
\caption{{\bf A} Fixation probability, $P_\text{fix}$, depending on selection strength, $s$, and variability $\sigma_1\!=\!\Delta$ according to Eq.~\eqref{eq:Langevin}. In the green regime where $P_\text{fix}>0.5$ the faster growing trait is favored while in the gray regime the less sensitive species has an advantage. The black line indicates the parabola $s=\Delta^2/2$, which is our prediction for $P_\text{fix}=0.5$.   The dashed gray line corresponds to a condition for the survival of a non-competing species derived by May~\cite{May:1973}. {\bf B} Comparison of our analytic calculations for the fixation probabilities (Eq.~\eqref{eq:fix_prob}) and the full two species model (Eq.\eqref{eq:Langevin}) for $s=\{0,0.5,2,5\}$ in $\{$red, violet, blue, green$\}$. Other parameters are $\nu_1\!=\!10$, $\sigma_2=0$, $\gamma=1$ and $K=100$ and $x_0=0.5$. \label{fig:A}}
\end{figure}

Even though environmental variability causes a drift term favoring the trait which is less sensitive to environmental changes~\cite{Frank:2011}, the interplay between drift and diffusion term has to be understood to predict the evolutionary outcome. Indeed, the environmental contribution to the drift caused by  $\sigma_S$ is intrinsically connected to the diffusion term. In other words whenever the fitness of a species is influenced by its sensitivity to environmental noise also neutral evolution is increased.Therefore we study the fixation probability, i.e. the probability that trait {\bf 1} fixates or trait {\bf 2} goes extinct. This quantity includes the effects of deterministic selection and random drift due to environmental and demographic noise and provides a complete picture of the evolutionary process. As detailedly shown in the SM, the fixation probability can be calculated by solving the backward FPE, $0=\mathcal{L}^\dag_{x_0}P_\text{fix}(x_0)$ associated with Eq.~\eqref{eq:FP} for the boundary conditions $P_\text{fix}(0)=0$ and $P_\text{fix}(1)=1$ where $x_0$ is the initial fraction of type {\bf 1}. The operator $\mathcal{L}^\dag$ denotes the adjoint of $\mathcal{L}$ defined in Eq.~\eqref{eq:FP}. The solution of the backward FPE is given by,
\begin{align}
P_\text{fix}(x_0)=\!\frac{\!1\!-\!\exp\!\left\{{\zeta\! \left[\text{Tanh}^{-1}\alpha \!+\!\text{Tanh}^{-1}\alpha (2 x_0-1)\right]}\right\}}{1-\exp{\left\{2\zeta\, \text{Tanh}^{-1}\alpha\right\}}},
\label{eq:fix_prob}
\end{align}
with  $\alpha\equiv \beta/\sqrt{8+\beta^2}$ and $\zeta\equiv 2K \left(\sigma_1^2-\sigma_2^2-2 s\right)/(\beta\gamma \sqrt{8+\beta^2})$ where $\beta=\sqrt{K(\sigma_1^2-2\epsilon\sigma_1\sigma_2+\sigma_2^2)/\gamma}$. 

In Fig.~\ref{fig:A} we show the fixation probability  for different values of $s$ and $\sigma_1=\Delta$ ($\sigma_2=0$) and compare it to our analytic calculations. Results are obtained by  numerical solution of Eq.~\eqref{eq:Langevin}, i.e. before any approximation or simplification has been performed. One can clearly distinguish two distinct regions where one of the two species is is predominant: in the gray area the smaller variability dominates while in the green regime the faster growing species prevails. To test our hypothesis that the drift term and the resulting stable fixed point is responsible for this behavior we additionally plot the condition for a stable fixed point at $x^*=0.5$ (solid black line) given by $\Delta=\sqrt{2s}$. This line indicates the point in phase space where both species are equally likely to survive and therefore equally fit. As expected the line separates the two regimes where either the faster growing or the less sensitive species survives.
In panel {\bf B} of Fig.~\ref{fig:A}, the fixation probability depending on $\Delta$ is compared to the analytic solution (Eq.~\eqref{eq:fix_prob}) for four values of $s=\{0,0.5,2,5\}$. Both show very nice agreement proving that the approximations which were made to derive the one-dimensional FPE, Eq.~\eqref{eq:FP}, are valid and confirming conclusions drawn from it. 

The general case of both species having variable birth rates yields analogous results: a selection advantage for the species with less sensitivity on the environment. Importantly, the selection disadvantage due to environmental noise does not scale with the carrying capacity as most demographic fluctuation effects do, i.e. the mechanism is effective irrespectively of the population size. When investigating only one species, the condition $\Delta^2/2>\nu$ for a quick extinction of this species was derived in Ref.~\cite{May:1973} and is indicated in Fig.~\ref{fig:A}~{\bf A} by a dashed gray line.  Our results manifestly go beyond that limit and a substantial disadvantage for the more variable species is present even when its variability is small enough to  ensure survival in a scenario without competition.

An alternative interpretation of our results on the fixation probability is as follows. As mentioned above, the parameters $\sigma$ or $\Delta$ depend on two factors: the sensitivity of a trait's growth rate on environmental conditions and the strength of the environmental fluctuations themselves. For a given sensitivity of an individual on the environment, the ordinate $\Delta$ in Fig.~\ref{fig:A} then corresponds to the strength of environmental fluctuations. While for weakly fluctuating environments a growth advantage is more beneficial, the situation is different for  strong environmental variations. Then it is more advantageous to minimize the sensitivity to those variations rather than to optimize the growth rate. Interestingly one can construe this result in the context of game theory: decreasing the sensitivity to environmental changes also means to optimize the worst-case-scenario outcome because the average birth rate is the least reduced when the variability is small. In game theory, this corresponds to the MaxiMin strategy which was  shown to be very successful in many fields as finance, economy or behavioral psychology~\cite{Neumann:1944,Owen:1995}. In the field of evolutionary dynamics another example of a MaxiMin strategy is bacterial chemotaxis, where it was proposed that bacteria track chemoattractants trying  to maximize their minimal uptake~\cite{Celani2009}.
\subsection*{Neutral Evolution}

Beside contributing to the fitness, environmental variability also influences fixation probability and time in the case of  \emph{neutral evolution}, i.e. $\nu_1\!=\!\nu_2$ and $\sigma_1\!=\!\sigma_2=\sigma$. Such analysis is of great interest, as evolution is often studied by investigating how neutral mutations evolve over time. In recent years fast-sequencing techniques made huge amounts of data available~\cite{Shendure:2008} which is now analyzed. Often it is interpreted by comparison to evolutionary models as  for example the Moran or Wright-Fisher model~\cite{Blythe:2007}. In particular, the population size of ideal population which produces similar results in the toy model, the effective population size, is often inferred and used as characteristic quantity~\cite{Charlesworth:2009ph}. Interestingly it was observed that often thereby calculated population sizes are much smaller than the census population size. This observation is still present when one corrects for factors like a finite fertility span or the sex ratio~\cite{Charlesworth:2009ph}. In the following, we want to demonstrated that in addition to already know factors also  environmental noise can account for a discrepancy between effective and census population size.

While the correlation parameter does not qualitatively influence results discussed so far, it plays an important role for neutral evolution. Interestingly, for fully correlated noise $\epsilon=1$, i.e. when both species are subjected to the same fluctuations meaning that even though the growth rates are variable their values are always the same, the resulting dynamics is exactly the same as in the case without environmental noise.  Therefore, extinctions are solely driven by demographic fluctuations and well-known results apply~\cite{Kimura}. Mathematically, this is reflected in Eq.~\eqref{eq:FP}, which for fully correlated noise, $\epsilon=1$,  is the same as for $\sigma_1=\sigma_2=0$.  In contrast, for all other values of $\epsilon$, including uncoupled noise $\epsilon=0$, the dynamics differs in two major respects. First, the drift term $-\sigma^2\left(1-\epsilon\right)\partial_x(1-2x)x(1-x)P(x)$ does not vanish and corresponds to a stable fixed point at $x^*=0.5$ pushing the system to coexistence. Second, the diffusion term consists of demographic $\frac{\gamma}{K}x(1-x)$ and environmental fluctuations $\left(1-\epsilon\right)\sigma^2x^2(1-x)^2$ leading to a larger randomness in the evolutionary behavior.  As the Darwinian drift suppresses extinction events while a larger diffusion term favors them, a more detailed analysis is required to grasp the evolutionary outcome. 

\begin{figure}[t]
\centering
\includegraphics[width=\columnwidth]{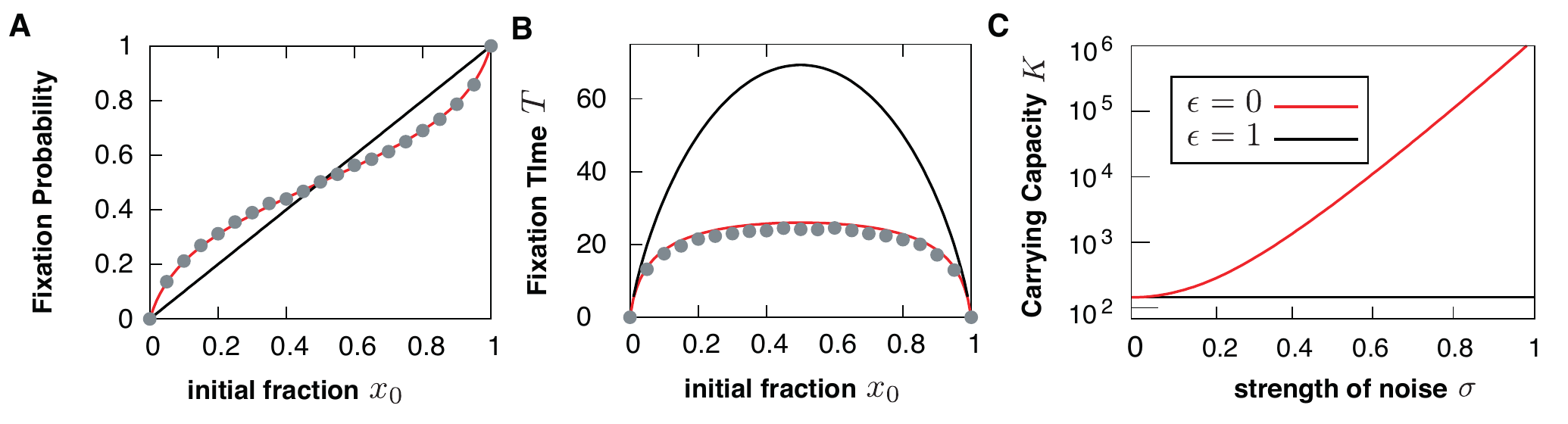}
\caption{Fixation probability (panel {\bf A}) and  time (panel {\bf B}) in the neutral case. Solid lines indicate analytical results for the two typical cases of perfectly correlated and uncorrelated noises $\epsilon=1$ and $\epsilon=0$.
Parameters are\,: $\nu_1=\nu_2=10$, $K=100$, $\sigma_1\!=\!\sigma_2\!=\!0.5$ and $\gamma=1$. Dots are simulations of the IBM and show good agreement with the analytic results. Additional parameters are $m=1,~\phi_1\!=\!\phi_2\!=\!10$, $  \omega_1\!=\!\omega_2=5$, $\tau\!=\!0.01$, $\langle E\rangle\!=\!0$, $\alpha_1=\alpha_2=1$, and $\text{Var}[E]\!=\!100$. {\bf C}
Reduction of the effective population size due to environmental noise in the neutral case for $x_0=0.5$. Using \eqref{eq:fix_time}, we plot the values of the carrying capacity $K$ and noise 
$\sigma$ that lead to the fixation time $T=100$ for $\gamma=1$. The black line corresponds to $\tilde\sigma=0$ (see Eq.~\eqref{eq:fix_time}), i.e. either perfectly correlated noise or no environmental noise. In the presence of environmental  noise, the values of $K$  are systematically higher and  increase  several orders of magnitude even for moderate noise levels.
\label{fig:2}}
\end{figure}

Let us first consider the fixation probability [cf. Eq.~\eqref{eq:fix_prob}],  see Fig~\ref{fig:2} {\bf A}. While for the standard situation of no environmental noise (or fully correlated noise) a linear dependence of the fixation probability on the initial fraction of a trait, $x_0$ is observed, $P_\text{fix}^{\epsilon=1}=x_0$, [ Fig~\ref{fig:2} {\bf A} black line] for $\epsilon<1$ the situation is more complicated [red line]. Due to stable fixed point at $x^*=0.5$, a S-shape arises. This means that the fixation probability for both species depends less on the initial fraction than in the standard case and both species are more equally likely to fixate.

While the behavior of the fixation probability is mainly due to the stable fixed point, the situation is more intricate for the extinction or fixation time which is another important quantity to describe evolutionary processes. As mentioned above, a coexistence fixed point is expected to increase the extinction time while a larger random drift decreases it. To ultimately understand the influence of environmental fluctuations, we therefore calculated the extinction time, $T(x_0)$, analytically. It also obeys a backward FPE, $-1=\mathcal{L}^{\dag}_{x_0}T(x_0)$, which can be solved employing the boundary conditions $T(0)=T(1)=0$, see SM\,:
\begin{align}
T=\frac{1}{C\tilde{\sigma}^2}\left[
\ln\frac{1-\Gamma_{\!+}(1-x_0)}{1-\Gamma_{\!-}(1-x_0)}\ln(1-x_0)+\ln\frac{1-\Gamma_{\!+}x_0}{1-\Gamma_{\!-}x_0}\ln x_0 +F_{\Gamma_+}(x_0)-F_{\Gamma_-}(x_0)\right]\,,
\label{eq:fix_time}
\end{align}
where $\tilde{\sigma}=(1-\epsilon)\sigma$, $C=\sqrt{1+4\gamma/(K\tilde\sigma^2)}$, $\Gamma_\pm\!=\!2/(1\pm C)$,  the function $F_{\Gamma}(x)\equiv \text{Li}_2(\Gamma(1\!-\!x))+\text{Li}_2(\Gamma x)-\text{Li}_2(\Gamma)$ and $\text{Li}_n$ is the polylogarithm.

 The result for not fully correlated noise ($\epsilon<1$)  differs again from the non-fluctuating/fully correlated scenario, $T^{\epsilon\!=\!1}=-K/\gamma[x_0\ln(x_0)+(1-x_0)\ln(1-x_0)]$ (see Fig.~\ref{fig:2} {\bf B}). Fluctuating environments decrease the fixation times for all initial conditions despite the stable fixed point at $x^*=0.5$. In other words, the extinction time is more strongly influenced by the larger neutral drift than by the stable fixed point. This has a crucial consequence: when measuring extinction times and comparing them to  models without environmental fluctuations, one can only explain small fixation times which are due to larger fluctuations by demographic noise. Therefore, demographic noise has to account for all randomness and the resulting effective population size is much smaller than the census population size. Indeed, such a behavior is often found when analyzing data~ \cite{Charlesworth:2009ph}. Fig.~\ref{fig:2} {\bf C} shows that conspicuous orders-of-magnitude reductions in the population size set in already at moderate levels of environmental noise, which could then account for differences between effective and census population size. In other words, as long as the strength of environmental noise and its correlation level are not known, the effective population size can only be interpreted as a lower bound for the census population size. In this context, environmental noise could also account for discrepancies between effective population sizes which are inferred with different methods for the same population. For instance environmental noise is not expected to change the population size determined via the Hardy-Weinberg equilibrium where the population size scales the sampling noise~\cite{Charlesworth:2009ph} while it does when dynamic quantities like coalescence times are employed.

\subsection*{Individual Based Model}

To further investigate the impact of variable environmental conditions, we introduce an Individual Based Model (IBM).  Such individual or agent based models serve as powerful tools to study evolutionary processes. They intrinsically include demographic noise as reproduction and death events are explicitly modeled. Additionally the IBM here serves as a proof of principle that linear multiplicative noise can be realistically expected when considering birth rates which depend on fluctuating environments. Since we model the environment and its fluctuations now explicitly, we can vary the environmental switching rate and thereby study the so far discussed phenomena beyond the white-noise limit, i.e. for environments which change slower. Even though, a particular choice for the IBM is made, we want to stress that the results presented above hold for any microscopic model whose macroscopic representation is given by Eq.~\eqref{eq:Langevin}, i.e. where the birth rate is subject to fluctuations thereby leading to linearly multiplicative noise. 

In the IBM each individual reproduces according to its experienced environments. The simplest version of the model is that only the current environment influences the birth rate but to show that our results are more rigorous we also include more realistic scenarios where an individual's environment history matters. To be more specific, the reproduction rate of an individual, $i$, at time $t$, depends a priori on the history of environmental conditions  experienced during its lifetime $t^i_\text{life}=[t_0^i,t]$, where $t_0^i$ is the time of birth of the $i$-th individual. This could for example account for the accumulated level of nutrients or detrimental substances that individuals are exposed to. Following~\cite{Melbinger:2010, Cremer:2011a}, our model is based on independent birth and death rates, i.e. a birth event is not necessarily coinciding with a death event allowing for variably population sizes. The birth rates now depend on the environment which is described by a parameter $E$. Without loss of generality, we assume that larger values of $E$ correspond to better environmental conditions. Depending on the correlation level of the noise, this parameter can be the same ($\epsilon=1$), completely independent ($\epsilon=0$) or correlated for different species. The number of environments experienced by an individual, $i$, is denoted as $M^i$ and their values are contained in a vector $\vec E^{i}=(E_1^{i},E_2^{i},...,E^i_{M^{i}})$. Environmental conditions change stochastically at rate $1/\tau$. At each switching event a new environment is drawn according to a  normal distribution, $p(E)$, with mean  $\langle E\rangle $ and variance $\text{Var}[E]$.  

The growth rate of a species depends on the previously experienced environments. Before considering that, let us first discuss how  the growth rate depends on a particular constant environment $E$. This quantity, the  \emph{instantaneous growth rate} $\lambda_S(E)$ is assumed to be a increasing function of the environment $E$ [see Fig.~\ref{fig:change_env} {\bf A}], i.e. in better environments individuals reproduce faster.  In particular, we consider the sigmoidal function\,:
  \begin{align}
 \lambda_S(E)=\phi_S+\omega_S\tanh\left(\alpha_S E/2\right)\,,
 \label{eq:lambda}
 \end{align}
 with $\phi_S$ the ordinate of the inflection point, $\omega_S\le \phi_S$ the maximal deviation from it, and $\alpha_S$ scaling the growth rate's sensitivity to the environment. 
 
Let us now consider changing environments and individuals whose current growth rate depends also on previously experienced environments. The reproduction rate $\Gamma^i_\text{repr,S}$ of an individual, $i$, of type $S$ now depends on the whole vector, $\vec E^{i}$.
For concreteness, we  assume that memory decays exponentially and that the rate is given by
\begin{align}
\Gamma^i_\text{repr,S}=\frac{1-m~~~}{1-m^{M^{i}}}\sum_{j=1}^{M^{i}}m^{j-1}\lambda(E^i_j),
\label{eq:Gamma}
\end{align} 
where the memory parameter $m\in[0;1]$ defines the influence of previously experienced environments upon an individual's growth rate. For $m=0$ only the current environment sets the growth rate $\Gamma^i_\text{repr,S}=\lambda(E^i_{M^i})$, and individuals do not memorize their past. In contrast, with increasing $m$ the previously experienced environments become more and more important. For the limit  $m\rightarrow1$ all experienced environments, $M^i$, have the same influence and the growth rate is given by the arithmetic mean of all experienced instantaneous growth rates $\Gamma^i_\text{repr,S}=\frac{1}{M^i}\sum_{k\leq M^i} \lambda^k_S( E_k^{i})$. We assume for simplicity 
that offsprings lose memory at the time of reproduction, independently of $m$. As in the Langevin model, limited resources restrict the maximal number of individuals in the population. Therefore, the death rates $\Gamma^i_\text{death,S}=\gamma N/K$, increase with the total population size $N$.
\subsubsection*{Mapping}
 To compare the results of the microscopic individual based model to the effective stochastic model, Eq.~\eqref{eq:Langevin}, the parameters of both models have to be mapped. In this section, we briefly explain how such a mapping can be obtained but results in the following sections can be understood without those details.. For simplicity let us consider the case $\langle E\rangle=0$ and a symmetric distribution $p(E)$ throughout the following discussion. Since death rates are constant, there is a direct correspondence between their value in the Langevin and the IBM. For birth rates and their STDs the situation is more intricate as we discuss hereafter. For the no-memory case ($m=0$) an exact mapping is obtained in the SM: For strong fluctuations, $\alpha_S^2\text{Var}[E]\gg 1$, the mean of the growth rate $\nu_S$ and the STD of the noise $\sigma_S$ in Eq.~\eqref{eq:Langevin} are given by\,:
\begin{align}
\nu_S(m=0)=\phi_S+\omega^2\tau,~~
\sigma_S(m=0)=\omega\sqrt{2\tau}.
 \label{eq:mapping0}
\end{align}
Note that the variability in the growth rate not only results in $\sigma_S>0$, but also influences the average reproduction rate $\nu_S$. While for $m=0$ such a variability increases $\nu_S$, the second term of $\nu_S$ is reduced while $m$ increases till it changes sign (see 
SM for details). For instance, for $m=1$ the growth rate is approximately $\phi_S-\omega^2\tau$. Hence, the more variable trait has a disadvantage in the average reproduction rate in addition to the effects discussed above. For $m=1$, the approximation $\sigma_S(m=1)\approx\omega_S\sqrt{\tau}$ holds. Dependencies in this expression are intuited as follows. The number of environmental changes an individual experiences until the memory resets is of the order $M\sim t_\text{life}/\tau$, where $t_\text{life}\propto 1/\nu_S$ is the typical time for an individual to reproduce or die. As environmental changes are independent random events, the variance of the reproduction rates \eqref{eq:Gamma} is $\propto\omega_S^2/M$. The expression for $\sigma_S(m=1)$ is finally obtained noting that correlations in the noise extend over times $\sim t_\text{life}$\,;  it follows then that the average reproduction rate $\nu_S$  drops out. 
\subsubsection*{Results Beyond the White-Noise Limit}

  \begin{figure}[t]
 \centering
 \includegraphics[width=\columnwidth]{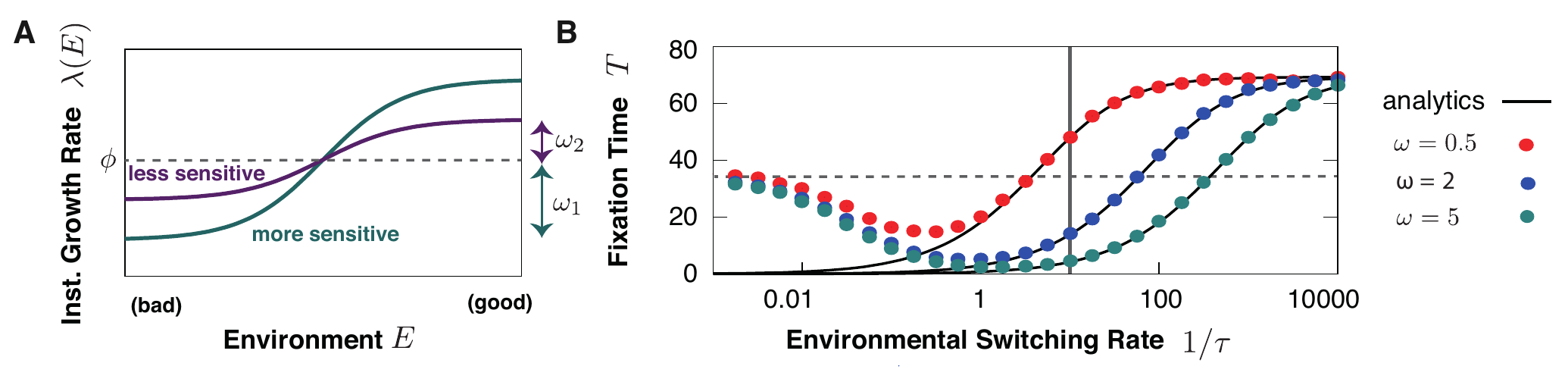}
 \caption{\label{fig:change_env}{\bf A} Illustration of the instantaneous growth rate depending on the environment. Both species have the same average reproduction rate $\phi_1=\phi_2=\phi$ but species {\bf 1} is more sensitive to environmental changes ($\omega_1>\omega_2$) {\bf B} Comparison of the IBM and the Langevin model. We show the fixation time for neutral evolution for $x_0=0.5$ {\it vs} the environmental switching rate $1/\tau$. Dots correspond to the IBM [$m=0$] for different values of $\omega_1\!=\!\omega_2=\{0.5,2,5\}$ in red, blue and green. Black lines are analytic solutions [Eq.~\eqref{eq:fix_time} with  Eq.~\eqref{eq:mapping0}]. For quickly fluctuating environments both results are in good agreement whilst for large $\tau$ the white noise approximation fails. The gray line corresponds to the timescale of reproduction. Other parameters are as in Fig.~\ref{fig:2}.
 }
 \end{figure}
 For a detailed comparison of the  IBM with the analytics derived in the first part of this paper, we simulate the IBM with a modified Gillespie algorithm updating reproduction rates after every environmental change~\cite{Gillespie}. As shown in Figs.~\ref{fig:2}{\bf A} and {\bf B}, results for fixation probability and time, are in excellent agreement with analytic solutions [Eqs.~\eqref{eq:fix_prob} and \eqref{eq:fix_time}]. In particular, the sigmoidal shape of the fixation probability is well reproduced by the IBM, supporting the existence and importance of linear multiplicative noise.

  Finally, the IBM enables us to study the environmental switching rate. This is of main interest as previous results were obtained using a white-noise approximation and strictly hold only for very rapidly fluctuating environments. In Fig.~\ref{fig:change_env} {\bf B}, the dependency on $\tau$ of the extinction time in the neutral case for $x_0=0.5$ is shown for different $\omega_S$; see SM for results with $s\neq0$. The black lines correspond to Eq.~\eqref{eq:fix_time} mapped according to Eqs.~\eqref{eq:mapping0} and dots are obtained by stochastic simulations of the IBM for $m=0$. For $\tau<1$ both models are in very good agreement. For larger $\tau$ switching is too slow to be well described by white noise. For $\tau\rightarrow\infty$ the environment never switches and as we choose a random distributed initial value of $E$ the average fixation time is given by the mean average extinction time: $\int_E p(E_1,E_2)T(E_1,E_2)$ where $p(E_1,E_2)$ is the joint probability distribution that species {\bf 1} experiences environment $E_1$ and species {\bf 2} experiences $E_2$ and $T(E_1,E_2)$ is the corresponding extinction time for constant environments. For large variances in the environmental distribution $\text{Var}[E]\rightarrow\infty$ the expression can be further simplified. The environmental conditions then only fluctuate between a good and a bad state. Therefore, only a few different outcomes are possible: either both species are in the the good (respectively bad) state and the extinction process is neutral or they are in different states and extinction is selection driven. As the latter, is much quicker than neutral extinction it can be neglected in the mean. The resulting extinction time is now approximately given by the probability for neutral evolution, i.e. that both individuals experience the same environment, which is given by $p_\text{neut}=0.5$ times the extinction time in the neutral case without environmental noise $T_\text{neut}=T^{\epsilon\!=\!1}=-K/\gamma[x_0\ln(x_0)+(1-x_0)\ln(1-x_0)]$. In Fig.~\ref{fig:change_env} {\bf B}, the dashed line marks the value of this approximation.

All in all, our analysis of the IBM beyond the white noise limit confirms that there is a broad parameter regime where environmental fluctuations play a crucial role for both neutral evolution and the fitness functions. For fluctuations up to the timescale of reproduction events (marked by the vertical gray line), the description introduced above is valid.  Nutrients and other metabolically important substances can vary on time scales quicker than reproduction. Therefore, we expect effects as discussed above to play a crucial role for evolutionary dynamics.
  
    \section*{Conclusion}
  
We quantitatively demonstrated  that environmental variability has crucial impact on evolutionary fitness.  Our results do not rely on details of microscopic models but are rather derived from a macroscopic model whose only key assumption is that the birth rate of individuals is not constant but fluctuates. This assumption automatically leads to linearly multiplicative noise which gives rise to all discussed effects.

First, we quantified the role of reduced sensitivity to environmental changes and determined how it increases the fitness. Even though the increase stems from noise, its amplitude does not drop as the total population size increases. Therefore, such a mechanism is effective also for very large populations, contrary to most other fluctuation-based effects. By  studying the interplay of the resulting evolutionary dynamics with random drift, we confirmed the importance of that fitness contribution and showed that  those contributions  are visible in a broad parameter regime. Importantly, even though  fluctuation driven the fitness contribution due to environmental noise is of the same order of magnitude than the contribution due to different growth rates and present for all population sizes. This finding is of great interest when thinking about whether a generalist or a specialist is evolutionary favored~\cite{Gilchrist:1995, Kassen:2002}. We can quantify that depending on the level of environmental noise two regimes are present: For strongly fluctuating environments it strongly pays off to be less sensitive to such changes (to be generalist) while for little environmental fluctuations is more beneficial to reproduce as quick as possible (be a specialist). 

 In addition,  we showed that the timescale of extinction in the neutral case is strongly affected by environmental noise. That provides a possible contribution to the reduction of effective population sizes, which are often found experimentally to be much smaller than the census population size. The reason is that environmental fluctuations increase the random drift that automatically results in smaller effective populations size, even if the source of the larger fluctuations is not demographic noise. Finally, we investigated individual based models that generate the linear multiplicative noise considered here. We thereby demonstrate that our description holds for fluctuation time scales up to the time scale of reproduction events.

As a future perspective, it will be of interest to study other forms of multiplicative noise in more detail, e.g.
a noise in the death rate $\gamma$ that would lead to a nonlinear dependency 
of the noise on the number of individuals.  Also the interplay between the noise-induced frequency dependence discussed here and the one resulting from payoff matrixes in standard evolutionary game theory is worth further investigation.
Finally, the question as to how fluctuation effects are influenced by reproduction rates that depend on time - a realistic model extension - remains open. With no environmental fluctuations, such rates would result in a smaller standard deviation of the expected time of reproduction and could potentially further increase the strength of the effects on fitness that we  presented here.

%

\section*{Acknowledgements}

We thank Jonas Cremer  for valuable  discussions and comments on the manuscript. AM acknowledges the German Academic Exchange Service (DAAD) for financial support.

\section*{Author contributions statement}
A.M and M.V. designed and performed the research and wrote the manuscript.

\section*{Additional information}
%
%

The authors declare no competing financial interests.


\newpage


\section*{Supplementary material}

# Supporting Information: The Impact of Environmental Fluctuations on Evolutionary Fitness Functions

Anna Melbinger and Massimo Vergassola

## ABSTRACT

In this supporting information, we provide more details to the calculations sketched in the main text. In particular, we derive the one component version of the Fokker-Planck equation and expressions for fixation probability and time. In addition, we show additional data for the individual based model. We especially focus on the derivation and discussion of the mapping depending on the memory of environments previously experienced by an individual.

In this Supporting Material, we provide more details on the calculations leading to the one-dimensional Fokker-Planck equation (FPE), Eq. (3) main text. Moreover, we present calculations for the average fixation time and the extinction probability. We discuss the mapping of the individual based model (IBM) to the Langevin model. For the no-memory limit we present an analytic derivation of the mapping. For other parameter values we give heuristic arguments that are supported by additional data. Finally, we present results for non-neutral evolution investigating the regime in which the white noise approximation is an adequate description for the evolutionary process.

## 1 Derivation of the one-dimensional Fokker-Planck equation

In this Section, we provide details how the Langevin equations,

$$\dot{N}_1 = N_1(\nu_1 - \gamma\tfrac{N}{K}) + \sqrt{N_1\left(\nu_1 + \gamma\tfrac{N}{K}\right)}\mu_1 + \sigma_1 N_1 \xi_1$$
$$\dot{N}_2 = N_2(\nu_2 - \gamma\tfrac{N}{K}) + \sqrt{N_2\left(\nu_2 + \gamma\tfrac{N}{K}\right)}\mu_2 + \sigma_2 N_2 \xi_2 \quad \text{(S1)}$$

can be transformed into the one-dimensional FPE presented in the main text. Both noise terms in Eq. (S1) are interpreted in the Ito sense[?] because they only should act as noise, i.e. enter the variances but do not affect the means, as it can be seen in the Fokker-Planck equation (S4) [or (3) main text].

From general results on stochastic processes (see[1]), it follows that the previous Langevin equation is associated to the following two-dimensional FPE :

$$\frac{\partial P(N_1,N_2,t)}{\partial t} = \varepsilon \partial^2_{1,2}\sigma_1\sigma_2 N_1 N_2 P - \sum_i \partial_i \left[\left(\nu_i - \gamma\tfrac{N}{K}\right)N_i P\right]$$
$$+ \frac{1}{2}\sum_i \partial_i^2 \left\{\left[(N_i\sigma_i)^2 + N_i\left(\nu_i + \gamma\tfrac{N}{K}\right)\right]P\right\}, \quad \text{(S2)}$$

where $\partial_i \equiv \partial_{N_i}$. The drift part is directly stemming from the non-fluctuating parts of the Langevin equations $N_S(\nu_S - \gamma N/K)$. Diffusion depends on the correlation level of the noises experienced by the two species. In particular, we have introduced the correlation coefficient $\varepsilon \equiv \langle \xi_1 \xi_2 \rangle / \sqrt{\langle \xi_1^2 \rangle \langle \xi_2^2 \rangle}$.

The case when the two noises are the same is given by $\varepsilon = 1$, when they are independent is $\varepsilon = 0$ and when they are anti-correlated is $\varepsilon = -1$.

To study the evolutionary dynamics associated to Eq. (S2), the relative abundances are the natural choice of variables. Therefore, we transform the absolute abundances $N_1$ and $N_2$ to $x = \frac{N_1}{N_1+N_2}$ and $N = N_1 + N_2$. To perform the change of variables, not only $N_1 = xN$ and $N_2 = (1-x)N$ have to be replaced, also the differential operators and the probability distribution have to be transformed. Ensuring that the latter is still normalized after change of variables, the Jacobian has to be introduced, $P(N_1,N_2) \to \frac{1}{N}P(x,N)$. The derivatives are given by, $\partial_{N_1} \to \frac{1-x}{N}\partial_x + \partial_N$ and $\partial_{N_2} \to -\frac{x}{N}\partial_x + \partial_N$.

After the change of variables, the FPE for $x$ and $N$ can now be further simplified exploiting the fact that the time scale of selection, $s = \nu_1 - \nu_2$, is much slower than the one of the population growth $\nu_1 x + \nu_2(1-x)$.[2] Therefore, we marginalize the FPE with respect to the total population size $N$. Thereby, the integrals $\int_0^\infty dN$ of $N$-derivative terms such as $\partial_N \bullet$ or $N\partial_N^2 \bullet = \partial_N(N\partial_N \bullet) - \partial_N \bullet$ vanish and the FPE simplifies to

$$\frac{\partial P(x,t)}{\partial t} = \partial_x\left\{\left[-s - \sigma_2^2(1-x) + \sigma_1^2 x + \varepsilon\sigma_1\sigma_2(1-2x)\right]Q\right\}$$
$$+ \partial_x\left(\frac{s}{N}Q\right)$$
$$+ \partial_x^2\left\{\left[\frac{\sigma_1^2 - 2\varepsilon\sigma_1\sigma_2 + \sigma_2^2}{2}x(1-x) + \frac{\gamma}{2K} + \frac{\nu_1 - sx}{2N}\right]Q\right\}, \quad \text{(S3)}$$

where $Q \equiv x(1-x)P(x,t)$. The drift term in the second line stemming from demographic fluctuations can be neglected as $N \gg 1$ holds. To finally arrive at the one-dimensional FPE employed in the main text, we compute the steady state population size $N^*$. As the deterministic differential equation for $N$ is given by

$$\dot{N} = N\left[x\nu_1 + (1-x)\nu_2 - \gamma\tfrac{N}{K}\right],$$

the fixed point for the populations size is $N^* = K/\gamma[\nu_1 x + \nu_2(1-x)]$. Employing that relation and the aforementioned

1

condition $s \ll v_1 x + v_2(1-x)$, the last term in Eq. (S3) can be simplified as $\frac{v_1 - sx}{2N} \approx \frac{\gamma}{2K}$, which finally leads to the one-dimensional FPE in the main text:

$$\frac{\partial P(x,t)}{\partial t} = \partial_x \left\{ \left[ -s - \sigma_2^2(1-x) + \sigma_1^2 x + \varepsilon \sigma_1 \sigma_2 (1-2x) \right] Q \right\}$$
$$+ \partial_x^2 \left\{ \left[ \frac{\sigma_1^2 - 2\varepsilon \sigma_1 \sigma_2 + \sigma_2^2}{2} x(1-x) + \frac{\gamma}{K} \right] Q \right\}, \quad \text{(S4)}$$

## 2 Fixation probability

In the following, we derive a general expression for the fixation probability. The calculations are analogous to the procedure for the neutral case described in the body of the paper. To determine the fixation probability the following backward equation has to be solved,

$$0 = x(1-x) \left\{ \left[ s + \sigma_2^2(1-x) - \sigma_1^2 x - \varepsilon \sigma_1 \sigma_2 (1-2x) \right] \partial_x \right.$$
$$\left. + \left[ \frac{\sigma_1^2 - 2\varepsilon \sigma_1 \sigma_2 + \sigma_2^2}{2} x(1-x) + \frac{\gamma}{K} \right] \partial_x^2 \right\} P_{\text{fix}}(x). \quad \text{(S5)}$$

Boundary conditions are $P_{\text{fix}}(0) = 0$ and $P_{\text{fix}}(1) = 1$. The solution to Eq. (S5) for the fixation probability is

$$P_{\text{fix}}(x) = \frac{1 - \exp\left\{\zeta \left[\text{Tanh}^{-1}\alpha + \text{Tanh}^{-1}\alpha(2x-1)\right]\right\}}{1 - \exp\left\{2\zeta \text{Tanh}^{-1}\alpha\right\}}, \quad \text{(S6)}$$

with

$$\beta = \sqrt{K(\sigma_1^2 - 2\varepsilon \sigma_1 \sigma_2 + \sigma_2^2)/\gamma}; \quad \alpha \equiv \frac{\beta}{\sqrt{8+\beta^2}};$$
$$\zeta \equiv \frac{2K(\sigma_1^2 - \sigma_2^2 - 2s)}{\beta \gamma \sqrt{8+\beta^2}}. \quad \text{(S7)}$$

The solution (S6) is obtained by integrating (S5) once, to find the gradient

$$\partial_x P_{\text{fix}}(x) = \text{const.} \frac{(1 + \alpha(2x-1))^{\zeta/2-1}}{(1 - \alpha(2x-1))^{\zeta/2+1}}. \quad \text{(S8)}$$

The expression (S8) is verified to be proportional to the derivative of $\left(\frac{1+\alpha(2x-1)}{1-\alpha(2x-1)}\right)^{\zeta/2}$ and boundary conditions are then imposed to fix the two constants of integration. The resulting expression is finally transformed into Eq. (S6) by using the elementary identity: $2\text{Tanh}^{-1}(x) = \log[(1+x)/(1-x)]$. It is verified that in the limit $\zeta \to 0$, one recovers the expression given in the main text.

All in all, the behavior we discussed in the main text is validated by analyzing the fixation probability: Both a higher growth rate and a smaller variability are beneficial for an individual.

## 3 Average time for fixation

### 3.1 Neutral case

The expression for the time of fixation in the neutral case that we presented in the body of the paper is derived as follows. The average time for fixation obeys the following backward equation,

$$\left\{ 1 - 2x + \left[ x(1-x) + \frac{\gamma}{K\tilde{\sigma}^2} \right] \partial_x \right\} \partial_x T(x) =$$
$$- \left( \tilde{\sigma}^2 x(1-x) \right)^{-1}, \quad \text{(S9)}$$

with $\tilde{\sigma} = (1-\varepsilon)\sigma$ the boundary conditions $T(0) = T(1) = 0$. Integrating Eq. (S9) and by variation of constants, we obtain:

$$\partial_x T(x) = \frac{1}{x(1-x) + \frac{\gamma}{K\tilde{\sigma}^2}} \left[ A + \frac{1}{\tilde{\sigma}^2} \ln\left(\frac{1-x}{x}\right) \right], \quad \text{(S10)}$$

where $A$ is a constant to be fixed by the boundary conditions. The integrals $\int_0^x$ of Eq. (S10) needed for $T(x)$ are performed by decomposing the rational function at the prefactor and using the formula:

$$\int \frac{\ln(a+bx)}{x} dx = \ln a \ln x - \text{Li}_2\left(-\frac{bx}{a}\right), \ a > 0, \quad \text{(S11)}$$

that follows from the very definition of the dilogarithm $\text{Li}_2(x) = -\int_0^x \ln(1-u)/u\, du$ (see[3]). The formula (S11) is used four times either directly (with a simple change of variables) or first integrating by parts to satisfy the condition $a > 0$ in (S11). The resulting expression is then transformed to the form given in the main text (which is the one given by Mathematica) by using the reflection property, $\text{Li}_2(x) + \text{Li}_2(1-x) = \text{Li}_2(1) - \ln x \ln(1-x)$, see.[3]

### 3.2 General case

In the general case when selection is present, the expression for the average fixation time cannot be found explicitly but is reducible to quadratures as follows. The fixation time obeys the backward equation (S5) with the left-hand side replaced by $-1$. Using the definitions (S7), we obtain

$$\left\{ 2\frac{K}{\gamma}(s + \sigma_2^2 - \varepsilon \sigma_1 \sigma_2) - 2\beta^2 x + \left[ \beta^2 x(1-x) + 2 \right] \partial_x \right\} \times$$
$$\partial_x T(x) = -\frac{2K}{\gamma x(1-x)}. \quad \text{(S12)}$$

Boundary conditions are $T(0) = T(1) = 0$. The homogeneous solution was already found following (S8) and reads

$$T_{hom}(x) = C_1 + C_2 \left(\frac{\chi_+(x)}{\chi_-(x)}\right)^{\zeta/2}, \quad \text{(S13)}$$

where $C_1$ and $C_2$ are constants and we defined

$$\chi_+(x) \equiv 1 + \alpha(2x-1), \quad \chi_-(x) \equiv 1 - \alpha(2x-1), \quad \text{(S14)}$$

to simplify notation. The non-homogeneous solution for the gradient of $T$ is obtained by varying the constant in (S8),

remarking that $\beta^2 x(1-x)+2 = \chi_+(x)\chi_-(x)(8+\beta^2)/4$ and integrating the resulting first-order differential equation to obtain

$$\partial_x T_{part}(x) = -\frac{K\chi_+(x)^{\zeta/2-1}}{2^{\zeta/2}\gamma(\zeta/2+1)} \times$$
$$\left[(\alpha-1)F_1\left(\frac{\zeta}{2}+1,\frac{\zeta}{2},1;\frac{\zeta}{2}+2;\frac{1}{2}\chi_-(x),\frac{\chi_-(x)}{1+\alpha}\right) + (\alpha+1)F_1\left(\frac{\zeta}{2}+1,\frac{\zeta}{2},1;\frac{\zeta}{2}+2;\frac{1}{2}\chi_-(x),\frac{\chi_-(x)}{1-\alpha}\right)\right]$$

where $F_1$ is the hypergeometric function of two variables.[4] The solution for $T$ involves the integral $\int_0^x \partial_y T_{part}(y)\,dy$ of the expression above (for which a closed form does not seem to be available), and the two constants in (S13) are fixed by

$$C_1 + C_2\left(\frac{\chi_+(0)}{\chi_-(0)}\right)^{\zeta/2} = 0$$
$$C_1 + C_2\left(\frac{\chi_+(1)}{\chi_-(1)}\right)^{\zeta/2} = -\int_0^1 \partial_y T_{part}(y)\,dy.$$

It is verified from the expression above or directly from the original equation (S12) that in the two limits $x \to 0$ and $x \to 1$ the solution behaves like in the neutral case, i.e. $-K/\gamma x \log x$ and $-K/\gamma(1-x)\log(1-x)$. Selection and the rest of the parameters affect of course the solution in the rest of the interval of definition $x \in [0,1]$.

## 4 Coexistence time

Depending on the position of the stable fixed point, coexistence between two species (one with a larger growth rate, one with a smaller variability, $v_1 > v_2$ and $\sigma_1 = \Delta$, $\sigma_2 = 0$) is possible. In this section we present some additional data demonstrating this. In Fig. S1, the extinction time which corresponds to the time of coexistence is shown depending on $\Delta$ is shown for different values of $s$. Dots correspond to solutions of Eqs. (S1) and black lines are numerical solutions of Eq. (S12). The extinction time has a maximum which exactly coincides with the parameter values of a fixed point $x^* = 0.5$. The dependence of this maximal extinction time on the selection strength $s$ is shown in Fig. S2.

## 5 Mapping individual-based models onto the Langevin dynamics

The aim of this Section is to show that individual-based models are described by the Langevin equations, Eqs. (S1), discussed in the main text and to analyze the mapping between the parameters of the two models.

The environmental conditions change stochastically at the rate $1/\tau$ and are distributed according to a distribution, $p(E)$, with mean $\langle E \rangle$ and variance Var[E]. The dependency of the instantaneous reproduction rate $\lambda_S(E)$ on $E$ is given by the sigmoidal function:

$$\lambda_S(E) = \phi_S + \omega_S \tanh\left(\frac{\alpha_S E}{2}\right), \quad (S15)$$

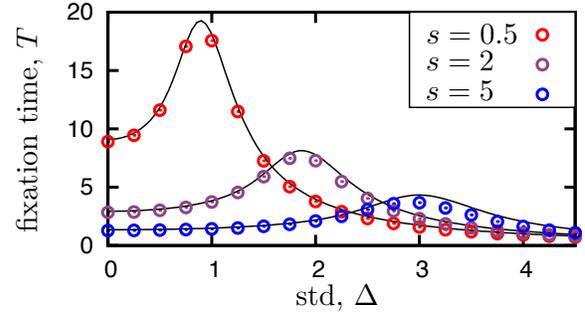

**Figure S1.** Extinction time depending on $\Delta$ for different values of the selection strength: $s = 0.5$ (red), $s = 2$ (violet) and $s = 5$ (blue). Dots are numerical solutions of the Langevin equations, Eq. (S1), and black lines are solutions of Eq. (S12).

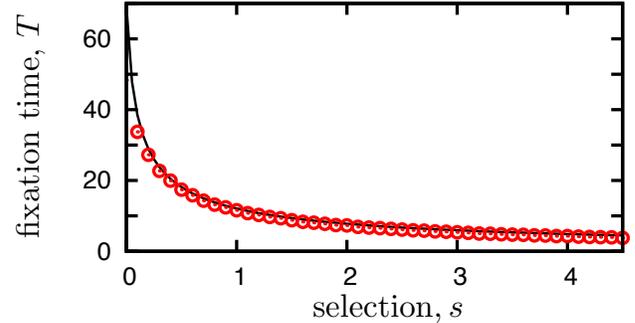

**Figure S2.** Extinction time for different values of $s$ and $\Delta = \sqrt{(2s)}$. This combination of $s$ and $\Delta$ corresponds to a stable fixed point at $x^* = 0.5$ and the maximal coexistence time for each value of $s$, see Fig. S1. As not only the selection strength but also the variability is increasing from left to right, the fixation time is a monotonically decreasing function of $s$.

which reduces to $\phi_S \pm \omega_S$ in the limit of large variances Var[E]. Birth rates are defined as,

$$\Gamma_{\text{repr},S}^i = \frac{1-m}{1-m^{M^i}} \sum_{j=1}^{M^i} m^{j-1} \lambda(E_j^i). \quad (S16)$$

In the no-memory limit $m = 0$, the growth rate is therefore given by the instantaneous growth rate $\lambda_S(E)$, while for $m \to 1$ the current growth rate is the arithmetic mean of all previously experienced environments. Death rates are given by $\Gamma_{\text{death},S}^i = \gamma N/K$.

### 5.1 No memory, $m = 0$

We discuss first the model without memory, where the memory parameter, $m$, is zero: Individuals reproduce with the instantaneous reproduction rates [Eq. (S15)], which reduce to $\phi_S \pm \omega_S$ in the limit of large environmental variance. We consider an interval of length $\delta t \gg \tau$ such that the probability for

an individual to reproduce or die is small, yet the total number of events occurring over the whole population ($\sim K \gg 1$) is large. Neglecting the standard demographic noise term,[5] the variation of the $S$-type population is given by

$$N_S(t+\delta t) \simeq N_S(t) + N_S(t)\left(\phi_S - \gamma\frac{N(t)}{K}\right)\delta t +$$
$$+ \omega_S \int_t^{t+\delta t} N_S(s)\hat{\sigma}(s)\,ds \quad (S17)$$

where $\hat{\sigma}(s)$ is the environmental Boolean random variable that takes values $\pm 1$ and switches with characteristic time $\tau$. The last term of Eq. (S17) is estimated as follows

$$\int_0^{\delta t} N_S(t+s)\hat{\sigma}(t+s)\,ds \simeq N_S(t)\mathscr{G}^e +$$
$$+ N_S(s)\omega_S \int_0^{\delta t} \hat{\sigma}(t+s)\,ds \int_0^s \hat{\sigma}(t+s')\,ds', \quad (S18)$$

where $\mathscr{G}^e$ is a Gaussian random variable having zero mean and variance

$$\text{Var}[\mathscr{G}^e] = \int_0^{\delta t} ds \int_0^{\delta t} ds' \langle \hat{\sigma}(s)\hat{\sigma}(s')\rangle = 2\tau\delta t. \quad (S19)$$

Here, we used that $\langle \hat{\sigma}(t)\hat{\sigma}(t')\rangle = e^{-|t-t'|/\tau}$ and $\delta t \gg \tau$. The second term in Eq. (S18) is evaluated at the order $\delta t$ using the same integral, Eq. (S19), and gives $N_S(t)\omega_S\tau\delta t$. Combining back all the terms, we conclude that the equation (S17) is equivalent to the Langevin equation (S1) with the mapping of the parameters

$$\nu_S = \phi_S + \omega_S^2\tau; \quad \sigma_S^2 = 2\omega_S^2\tau. \quad (S20)$$

Note that the standard demographic noise term in Eq. (S1) should a priori include the fluctuating environmental term $N_S(t)\omega_S\mathscr{G}^e$ in the sum of the rates. In fact, it can be safely ignored as $\phi_S N_S \delta t \gg \omega_S N_S \sqrt{2\tau\delta t}$ due to $\phi_S \geq \omega_S$ and $\delta t \gg \tau$.

Finally, the factor 2 appearing in $\sigma_S^2$ in (S20) depends on the Poisson statistics of the environmental fluctuations. If the duration is fixed and equal to $\tau$, Eq. (S19) becomes $\tau\delta t$. In that case, the corresponding mappings are $\nu_S = \phi_S + \omega_S^2\tau/2$ and $\sigma_S^2 = \omega_S^2\tau$. This is confirmed numerically in Fig. S3 where we show data for exponentially distributed (black) and fixed duration (red) environments. Solid lines are analytic solutions of the fixation time [Eq. (6) main text] employing the respective mappings.

### 5.2 Finite memory, $m > 0$

We now turn to the scenario where memory extends over several environmental conditions that an individual previously experienced. Whilst for $m = 0$ an exact analytic mapping can be found in the limit of small $\tau$, for finite memories the situation is more intricate. However, for the special case $m = 1$ the variability in the growth rate can be well approximated by the following argument. The reproduction rate $\Gamma_{\text{repr},S}^i$ of

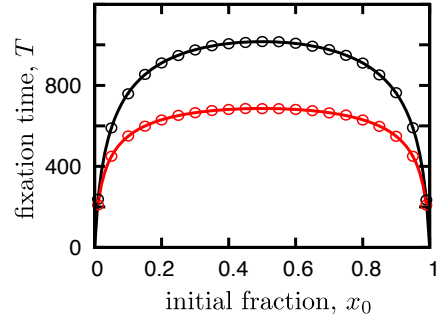

**Figure S3.** Comparison of data obtained by simulations in the neutral case with fixed (red) and exponentially distributed (black) environmental changes. Both sets of data agree with our analytic calculations, where we used the mappings $\sigma^2 = \omega^2\tau$ for fixed times of environmental changes and $\sigma^2 = \omega^2 2\tau$ for exponentially distributed switches. Thereby, the data confirms that the origin of the factor 2 in the mapping is solely the exponential distribution of the environmental changes. Parameters are $\phi_1 = \phi_2 = 1$ and $\omega_1 = \omega_2 = 0.9$, $\gamma = 1$, $K = 5000$, $\langle E \rangle = 0$, $\text{Var}[E] = 100$ and $\tau = 0.01$.

an individual, $i$, of type $S$ at time $t$ depends on the average of all instantaneous reproduction rates $\lambda_S(E)$ experienced by the individual:

$$\Gamma_{\text{repr},S}^i(t) = \frac{1}{M^i(t)} \sum_{k \leq M^i(t)} \lambda_S^k(E_k^i) = \phi_S + \tilde{\Gamma}_S^i(t). \quad (S21)$$

At the time of reproduction, we assume for simplicity that offspring looses its memory of past environments experienced by the progenitor.

We consider again a time interval of length $\delta t$ as in (S17). Fluctuations in the rates $\tilde{\Gamma}_S^i(t)$ decorrelate on timescales of the order of the lifetimes of individuals, $t_{\text{life}}$, which are much longer than $\tau$ and $\delta t$. Therefore on the $\delta t$ scale, noise is smooth, contrary to (S17). Conversely, timescales of several lifetimes are much smaller than those on which selection acts and much longer than the characteristic time of the noise. Therefore, to describe the dynamics of the fractions, the environmental noises are well approximated by a shortly correlated noise. An estimation of the amplitude of the noise is obtained by calculating the sum

$$\sigma_S^2 \sim \frac{1}{\tau} \sum_{\ell=-\infty}^{\infty} \sum_{i=1}^{N_S} \sum_{j=1}^{N_S} \langle \tilde{\Gamma}_S^i(t_k)\Delta t_k \tilde{\Gamma}_S^j(t_\ell)\Delta t_\ell\rangle.$$

The durations $\Delta t$ of the environmental intervals are independent for different $k$ and $\ell$ (and the contribution $k = \ell$ is negligible with respect to the rest of the sum) so that one can replace them by $\tau$. In addition, the symmetry in the indices of the intervals allows us to further simplify the expression

$$\sigma_S^2 \sim 2\tau \sum_{\ell=k}^{\infty} \sum_{i=1}^{N_S} \sum_{j=1}^{N_S} \langle \tilde{\Gamma}_S^i(t_k)\tilde{\Gamma}_S^j(t_\ell)\rangle. \quad (S22)$$

To compute the average in Eq. (S22) three different orders of events have to be distinguished: a) If the birth of the $j$-th individual was prior to the one of the $i$-th individual, then it follows from Eq (S21) that the quantity to be averaged is $\omega_S^2 M_k^i / (M_k^i M_\ell^j) = \omega_S^2 / M_\ell^j$, where $M_\ell^j$ is the number of environmental switches since the birth of the $i$-th individual up to time $t_k$. To derive Eq. (S22) we have used that the terms in the sum (S21) take independent values $\pm \omega_S$ with equal probability. Conversely, case (b) is when the birth of the $j$ individual is posterior to the one of the $i$-th individual. Then the quantity to be averaged is $\omega_S^2 (M_k^i - \delta b) / (M_k^i M_\ell^j)$, where $\delta b$ is the time between the birth of the $i$-th and the $j$-th individuals. Finally, in case (c) when $\delta b > M_k^i$, the correlation is zero as there is no overlap between the environmental fluctuations of the two individuals. Due to the Poissonian nature of the events, the number of switches since birth (back in the past) or before reproduction (forward in the future) have the same distribution $\exp(-M/\overline{M})/\overline{M}$ where $\overline{M} \sim t_{\text{life}}/\tau$ (its exact value does not affect the sequel). It follows that

$$\sigma_S^2 \sim \frac{2\omega_S^2 \tau}{\overline{M}^2} \left[ \int_0^\infty dt \int_0^\infty du \int_0^\infty dv\, e^{-(t+2u+v)/\overline{M}} \frac{1}{u+v+t} \right. \\ \left. + \int_0^\infty dt \int_0^\infty du \int_0^u dv\, e^{-(t+2u-v)/\overline{M}} \frac{u-v}{(u-v+t)u} \right]. \tag{S23}$$

The integral over $t$ is the continuous approximation of the sum over $\ell - k$ appearing in Eq. (S22) while the variables $u$ and $v$ refer to the variables $M_k^i$ and $\delta b$. The first and second term in the square parentheses of (S23) correspond to cases (a) and (b), respectively. By a series of change of variables and integrations by parts, it is shown that (S23) reduces to

$$\sigma_S^2 \sim \frac{2\omega_S^2 \tau}{\overline{M}^2} \int_0^\infty du\, e^{-u/\overline{M}} \int_0^u dv\, e^{-v/\overline{M}} = \omega_S^2 \tau. \tag{S24}$$

The validity of this approximation is confirmed for the neutral case in Fig. 2 in the main body of the paper where the fixation probability and time are compared to the analytic calculations employing Eq. (S24). Additional data for non-neutral evolution is presented in Fig. S4 where two species (one with finite variability, $\omega = 0.9$, one with vanishing variability) are analyzed. Analytic solutions for the fixation time and probability are fitted to simulation data. The best fit deviates less than 1% from Eq. (S24).

We now briefly discuss the dependence of $\sigma_S$ on the memory parameter $m$. In Fig. S5, the STD of the noise in the growth rate, $\sigma$, depending on the memory parameter, $m$, is analyzed. Results were obtained by simulating the neutral evolution case, measuring the fixation time and calculating $\sigma$ employing the analytic expression for the extinction time [Eq. 6 main text]. For $m = 0$ the thereby obtained value agrees nicely with the mapping introduced above indicated by the red dashed line [Eq. (S20)]. With increasing memory, $m$, the STD of the noise, $\sigma$, decreases monotonically and approaches Eq. (S24) for $m \to 1$.

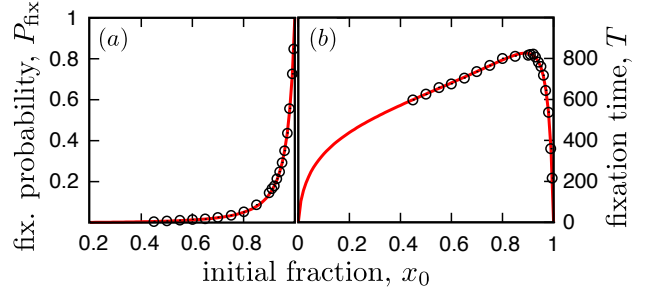

**Figure S4.** Fixation probability and time for non-neutral evolution with memory $m = 1$. Black dots correspond to simulation results of the IBM and red lines are analytic solutions. To obtain the latter we fitted Eq. (S6) and the solution of Eq. (S12) to the IBM. We used $\tilde{A} = s + \sigma_2^2 - \sigma_1 \sigma_2 \varepsilon$ and $\tilde{B} = \sigma_1^2 + 2\sigma_1 \sigma_2 \varepsilon + \sigma_2^2$ as fitting parameters and obtained $\tilde{A} = -3.04 \times 10^{-3}$ and $\tilde{B} = 8.25 \times 10^{-3}$. Other parameters are $\phi_1 = \phi_2 = 1$, $\omega_1 = 0.9$, $\omega_2 = 0$, $\tau = 1/100$, $\gamma = 1$, $K = 5000$, $\langle E \rangle = 0$ and $\text{Var}[E]$.

Let us now analyze the mapping of the average reproduction rate $v_S$. Importantly, this mapping is very sensitive to model details which we will exemplify in the following. As results for neutral species do not dependent on the average reproduction rate, we have to turn to the evolution of non-equal individuals to understand the mapping of the growth rates. In Fig. S4, we show the fixation probability for two species with the same $\phi_1 = \phi_2 = 1$ but only the first species has a variable reproduction rate $\omega_1 = 0.9$, $\omega_2 = 0$ for $m = 1$. Red lines correspond to a fit with $s = v_1 - v_2 \approx -0.0030$ and agree perfectly with simulation results. In other words, the first species does not only have a disadvantage due its sensitivity on environmental changes, $\sigma_1 > \sigma_2$, but also has a smaller average growth rate. To study this effect in more detail, let us now analyze the fixation probability dependence on the memory parameter $m$, see Fig. S6 panel (a). Black dots correspond to the standard IBM (if not mentioned otherwise our discussion applies to this data), red dots to a slightly changed model which is going to be introduced in the following. For $m = 0$ both species are equally likely to fixate as the growth advantage of the more variable species $v_1 = \phi + \omega_1^2 \tau > v_2 = \phi$ exactly compensates for its disadvantage due to the STD of the noise $\sigma_1 > \sigma_2$. For increasing values of $m$ first the more variable ($m < 0.7$) later the less variable species is favored ($m > 0.7$). Whether this behavior is caused by the STD of the noise or differences in the mean reproduction rates is not obvious as the influence of both fitness contributions is of comparable strength. Therefore, we estimate the selection coefficient, $s = v_1 - v_2$, from the fixation probability data, see Fig. S6 panel (b). This is achieved by assuming that the variability of species **1** with $\omega_1 = 0$ is zero ($\sigma_1 = 0$) and that the variability of species **2** with $\omega_2 = 8.237$ is the same as in the neutral evolution scenario and thereby given by the data presented in Fig S5.

**5/7**

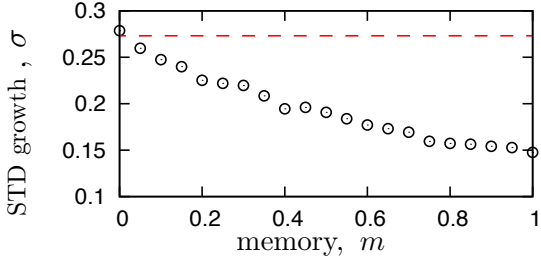

**Figure S5.** Numerical estimation of the STD of the noise dependence on the memory parameter $m$. Extinction times in the neutral scenario were measured. By evaluating the inverse of the fixation time function [Eq. (6) main text] the STD of the growth was calculated. For $m=0$ the result agrees with the mapping, cf. Eq (S20), indicated by the red dashed line. For larger values of $m$ the variability is reduced. Parameters are $\phi_1 = \phi_2 = 10$, $\omega_1 = \omega_2 = 8.237$, $\tau = 1/500$, $\gamma = 1$, $K = 100$, $\langle E \rangle = 0$ and $\text{Var}[E]$.

Note that this approximation might neglect some higher noise correlations arising due to the coupling of both species via the carrying capacity. For $m = 0$ the thereby obtained value of $s$ agrees well with our analytic results, cf. Eq. (S20). With increasing $m$ the growth rate of the more variable species is decreased till the selection coefficients becomes negative effectively favoring the more variable species. However the decrease of the selection coefficient with $m$ is smaller than the reduction of $\sigma$ shown in Fig. S5. Hence, for small $m$ the more variable species is favored as its advantage due to a larger average reproduction rate is larger than its disadvantage due to its sensitivity on the environment. This advantage in the growth rate is more sensitive to details in the IBM in comparison to the variability discussed in the main text for the Langevin equation Eq. (1).We are going to illustrate this by analyzing a slightly modified version of the IBM. But before doing so, we present an intuitive argument explaining one factor influencing the average growth rate: When an individual is born it experiences the current environment shorter than the average length of an environment. However, the model weights all experienced environments equally, see Eq. (7) in the main text. As the first experienced environment is more likely to be a good environment [more reproduction events happen during more beneficial environments], higher growth rates have a larger weight in the average and the average growth rate of the variable species is effectively increased. To obtain a description including this factor, it would be best to perform a time average over all previously experienced environments. Unfortunately, such a procedure is computationally very expensive. We therefore, test our explanation for the bias by not including the very first environment, the one in which an individual is born, in the averaged reproduction rate. The red dots in Fig. S6 corresponds to simulation results for this modified model. Even though all parameters are the same and for $\tau = 1/500$ and $\phi \approx 10$ an individual experiences in

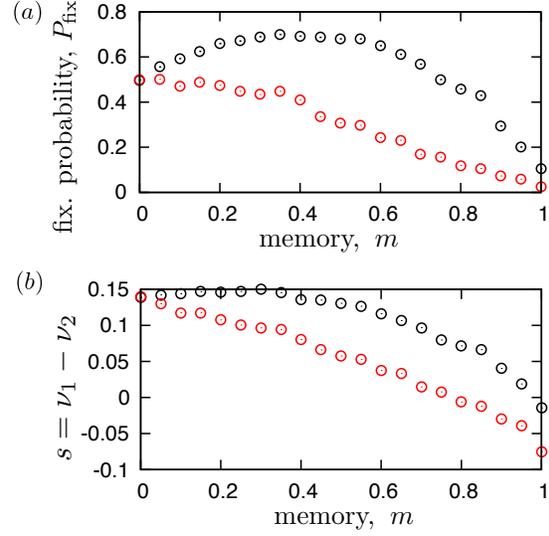

**Figure S6.** (a) Fixation probability and (b) selection coefficient depending on the memory parameter $m$. The first species' growth rate depends on the environment while the second one's is constant. Black dots correspond to the IBM introduced in the main body of the paper, while red dots represent a model modification not memorizing the first environment (the one in which an individual is born; for details see text). While the fixation probability is a direct simulation result, the selection coefficient $s$ is inferred from it using the additional data presented in Fig. S5. Parameters are $\phi_1 = \phi_2 = 10$, $\omega_1 = 8.237$, $\omega_2 = 0$, $\tau = 1/500$, $\gamma = 1$, $K = 100$, $\langle E \rangle = 0$ and $\text{Var}[E]$.

average 25 different environments, the small modification of the model substantially changes the simulation results. While the modification almost has no impact on the STD of the noise, it alters the average reproduction rate. For instance the regime in which the more variable species is favored completely disappears, cf. red dots Fig. S6 where $P_{\text{fix}} \leq 0.5$ for all $m$. This example illustrates that on the first sight tiny details of an IBM might substantially influence the evolutionary outcome and that one should be cautious when drawing conclusions from them. Importantly, the mechanism discussed in the main text does not rely on specific assumptions of the microscopic models: A finite STD of the growth rate is always a disadvantage. It might be compensated for by a larger average reproduction rate but the same value of the growth rate without variability is always preferable.

# 6 DEPENDENCE ON THE SWITCHING RATE

In this Section, we present additional data for the non-neutral case. Fig. S7 shows the fixation probability depending on the environmental switching rate $1/\tau$. In particular, we investigate extinction for a species which is not sensitive to its environment ($\phi_1 = \phi = 10$, $\omega_1 = 0$) competing with a sen-

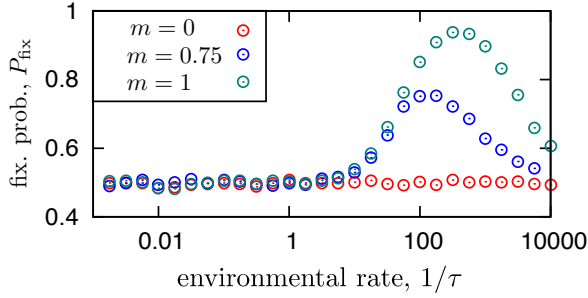

**Figure S7.** Dependence of the fixation probability on the environmental switching rate. Both species have the same $\phi_1 = \phi_2 = 10$, but only the second species is sensitive to the environment ($\omega_1 = 0$, $\omega_2 = 9$). For no memory ($m = 0$), both species are equally likely to fixate, as the advantage in the average growth rate of species **2** exactly compensates for its disadvantage due to its sensitivity on the environment. For $m > 0$ those two effects do not cancel out anymore and the second species is favored. Other parameters are $\gamma = 1$, $K = 100$, $\alpha = 1$, $\langle E \rangle = 0$ and $\text{Var}[E] = 100$.

sitive species ($\phi_2 = \phi = 10$, $\omega_2 = 9$) for different values of $m$. In the case of no memory $m = 0$ both species are equally likely to fixate as the advantage in the average reproduction rate $v_2 = \phi + \omega_2 \tau$ exactly compensates for the disadvantage due to the STD of the noise in the growth rate [see Eq. (S20)]. For larger values of the memory parameter, a bias favoring the species with $\omega = 0$ is present (the exact value of the fixation probability depends on mapping details as discussed above). Importantly, the bias is not only present for very quickly fluctuating environments, but already emerges if reproduction events happen on a time scale comparable to $\tau$. This supports the conclusion, that we were already drawing in the body of this paper when discussing Fig. 4: the white noise approximation is an adequate description for such an evolutionary process in that parameter regime.



\end{document}